\documentclass[
twocolumn,
]{ceurart}

\usepackage{listings}
\usepackage{mathptmx}
\usepackage{xcolor}
\usepackage{xspace}
\usepackage{tikz}
\usetikzlibrary{shapes,backgrounds}
\usepackage[percent]{overpic}
\usepackage{pict2e}
\usepackage{amsmath}
\usepackage{cleveref}
\usepackage{wrapfig}
\usepackage[font=footnotesize,labelfont=bf]{caption}

\newtheorem{scenario}{Scenario}

\newtheorem{example}{Example}

\newcommand{\set}[1]{\ensuremath{\mathbf{#1}}\xspace}
\newcommand{\func}[1]{\ensuremath{\textsc{#1}}\xspace}
\newcommand{\param}[1]{\ensuremath{p_{#1}}\xspace}

\newcommand{\scen}[1]{\textsf{#1}}

\newcommand{\scenarios}[0]{\set{Scenarios}\xspace}
\newcommand{\parameters}[0]{\set{Params}\xspace}
\newcommand{\press}[0]{\func{Press}}
\newcommand{\setting}[0]{\func{Setting}}
\newcommand{\problem}[0]{\func{Problem}}
\newcommand{\scale}[0]{\set{Scale}}
\newcommand{\action}[0]{\func{Action}}
\newcommand{\response}[0]{\func{Response}}
\newcommand{\justification}[0]{\func{Justification}}
\newcommand{\category}[0]{\func{Category}}
\newcommand{\sound}[0]{\func{soundj}}
\begin{document}

\copyrightyear{2022}
\copyrightclause{Copyright for this paper by its authors.
  Use permitted under Creative Commons License Attribution 4.0
  International (CC BY 4.0).}

\conference{Ital-IA 2024: 4th National Conference on Artificial Intelligence, organized by CINI, May 29-30, 2024, Naples, Italy}

\title{On Representing Humans' Soft-Ethics Preferences As Dispositions}
\author[1]{Donatella Donati}[%
%orcid=0000-XXXX-XXXX-XXXX,
email=donatella.donati@univaq.it,
]
\cormark[1]
\address[1]{University of L'Aquila (UNIVAQ), L'Aquila, Italy}

\author[2]{Ziba Assadi}[%
%orcid=0000-XXXX-XXXX-XXXX,
email=ziba.assadi@gssi.it,
]
\address[2]{Gran Sasso Science Institute (GSSI), L'Aquila, Italy}

\author[1]{Simone Gozzano}[%
%orcid=0000-XXXX-XXXX-XXXX,
email=simone.gozzano@univaq.it,
]
%\address[4]{Organization, via Xxxx, Naples, 80125, Italy}

\author[2]{Paola Inverardi}[%
%orcid=0000-XXXX-XXXX-XXXX,
email=paola.inverardi@univaq.it,
]
%\cormark[1]
%\address[1]{Organization, via Xxxx, Naples, 80125, Italy}
%\address[2]{Organization,  Institution, via Xxxx, Naples, 80125, Italy}

\author[2]{Nicolas Troquard}[%
%orcid=0000-XXXX-XXXX-XXXX,
email=nicolas.troquard@gssi.it,
]
%\cormark[1]
%\address[1]{Organization, via Xxxx, Naples, 80125, Italy}
%\address[2]{Organization,  Institution, via Xxxx, Naples, 80125, Italy}

\cortext[1]{Corresponding author.}
%\fntext[1]{These authors contributed equally.}

\begin{abstract}
The aim of this paper is to represent humans' soft-ethical preferences by means of dispositional properties. We begin by examining real-life situations, termed as scenarios, that involve ethical dilemmas. Users engage with these scenarios, making decisions on how to act and providing justifications for their choices.
We adopt a dispositional approach to represent these scenarios and the interaction with the users. Dispositions are properties that are instantiated by any kind of entity and that may manifest if properly triggered. In particular, the dispositional properties we are interested in are the ethical and behavioural ones. The approach will be described by means of examples. The ultimate goal is to implement the results of this work into a software exoskeleton solution aimed at augmenting human capabilities by preserving their soft-ethical preferences in interactions with autonomous systems.
\end{abstract}


\maketitle
\section{Introduction}

The constant growth of interaction between human and artificial agents poses ethical challenges for our society. The autonomy that intelligent systems are increasingly acquiring allows human agents to delegate tasks and decisions to them. This delegation is, \emph{prima facie}, very convenient. Nevertheless, it deprives human beings of one of their most defining ethical aspects: their autonomy.

To contrast this situation, approaches that try to empower humans in their interactions with autonomous machines are sought. In this direction, we are interested in building personalised software solutions that allow an ethical mediation between human beings and automatic systems. That is, we want individuals' moral and behavioural preferences to be respected in the course of interactions that have moral significance. We are therefore in the domain of \emph{soft ethics}. Clearly, respect of the norms and accepted procedures is taken for granted and absorbed in the so-called ``hard ethics''. Hard ethics is what may contribute to making or shaping the law. % \cite{Floridi} 
To make the difference between hard ethics and soft ethics clearer, consider this quote from  \cite{Floridi}:
\begin{quote}
Soft ethics covers the same normative ground as hard ethics, but it does so by considering what ought and ought not to be done over and above the existing regulation, not against it, or despite its scope, or to change it, or to by-pass it (e.g.\ in terms of self-regulation). In other words, soft ethics is post-compliance ethics: in this case, `ought implies may'.
\end{quote}

It is therefore crucial to collect and represent individual soft ethics. In \cite{DBLP:conf/hhai/AlfieriDGGS23} it has been shown that it is possible to collect excerpts of people's moral and behavioural preferences from their responses to a questionnaire. Roughly, they developed a questionnaire composed of thirteen morally-loaded scenarios describing a context that involves a moral decisions to make. The user is then asked whether they would or would not undertake a certain action in that given context, and to justify their reply by assigning a value from 1 to 5 to four different parameters. 

\paragraph{A dispositional and behaviourist approach.}

In this paper, starting from the questionnaire, we aim at constructing a tentative model showing how this users' feedback can help in capturing users' soft ethics.
The model we propose represents individual soft ethics as \emph{dispositions} that, as explained in the following section, are well suited to capture the contextual nature of soft ethics.
Dispositionality can be acquired and probed through experience \cite{dispositional-modality}. 
This is analogous to the \emph{behaviourist} approach to learning agents' utilities in decision theory, where preferences are revealed by one's choices \cite{peterson}: an agent prefers $x$ to $y$ if and only if they choose $x$ over $y$ whenever given the opportunity.
In our study, the questionnaire is the probing method to elicit moral dispositions.
Through experience, one can build the ethical profile of an agent. This moral profile would be akin to a repertoire of (dispositional) rules indicating what action the agent would tend to take in a given context.

\paragraph{Outline.} We provide an overview of what dispositions are in \Cref{sec:dispositions}. In \Cref{sec:questionnaire-scenarios} we present the questionnaire of \cite{DBLP:conf/hhai/AlfieriDGGS23} and a clear identification of the pieces of information in the scenarios and the human agents' feedback. In \Cref{sec:consistency}, we specify what we may call a `moral oracle' which is used as a step for eliciting soft-ethics preferences from the existing questionnaire and feedback. The instrumental role delegated to this oracle motivates the future work, which is presented in a conclusion in \Cref{sec:outlook}.

\section{Dispositions}
\label{sec:dispositions}

Dispositionalism is a philosophical theory of properties. According to this theory, properties are potentialities of the objects that instantiate them: e.g., the fragility of glass, the solubility of a sugar cube, and the bravery of an individual. Fragility, solubility and bravery are potentialities that dispose the entities instantiating them to exhibit particular behaviours under specific circumstances. The glass is disposed to break if dropped on a hard surface, the sugar cube is disposed to dissolve if immersed in a cup of hot tea, and the courageous person is disposed to face challenges in a dangerous situation. Dispositional properties are modal in nature, which means that they individuate \emph{potential} behaviours of the entities possessing them, that is, what those entities \emph{could} do within a given context. We can summarise all this with two claims  that represent what Vetter calls ``standard conception of dispositions''; in her own words \cite{Vetter2015}:
\begin{enumerate}
    \item A disposition is individuated by the pair of its stimulus condition and its manifestation (or, if it is a multi-track disposition, by several such pairs): it is a disposition to $M$ when $S$ (or a disposition to $M1$ when $S1$, to $M2$ when $S2$, etc., if it is a mutli-track disposition).
    \item Its modal nature is, in some way or another, linked to or best characterised (to a first approximation) by a counterfactual conditional ``if $x$ were $S$, $x$ would $M$'' (or if it is a multi-track disposition, by several such conditionals).
\end{enumerate}

Let us clarify with an example: the courage of the individual (disposition $D$) is individuated by the pair of its stimulus condition that is the dangerous situation ($S$) and its manifestation that is the facing of the challenges by the individual ($M$). The relation between the disposition, the stimulus and the manifestation can be, roughly, individuated by the following counterfactual conditional: ``if the courageous individual were placed in a dangerous situation, the courageous individual would face the challenges.''

Another tenet of dispositionalism is that dispositions are \emph{gradable} properties:
a thin glass is more fragile that a sturdy vase, gasoline is more flammable than wood, some people are more courageous than others, etc. The dimension \emph{per se} is what Vetter calls potentiality, that is the fact something may manifest shattering, combustion or facing challenges. Depending on the context, types of objects differentiate themselves, for instance, in their fragility. Fragility is a determinate that is manifesting more or less easily the breakability (a neutral potentiality) of objects. This could be understood as a variation along the determinable-determinate dimension. There are many ways in which something may manifest breakability or combustion, so such manifestation could be further determined in nuances of fragility and inflammability. The specific way in which, for example, gasoline manifests combustion is its determinate way, which is different from the way in which wood manifests combustion.

There are various theories about dispositions and different versions of those theories. However, for the project at hand, using this standard conception is enough. This minimal version of dispositionalism is already helpful in representing the soft-ethical preferences of individuals.
An attempt to connect ethics and dispositions has been made before in \cite{dispositions-ethics}.

The questionnaire is the method to elicit the soft-ethical dispositions of human agents. The next section presents it in detail.

\section{A formal analysis of the questionnaire's scenarios}
\label{sec:questionnaire-scenarios}

This section presents the questionnaire and the scenarios that compose it, what human agents' feedback about the scenarios is made of, and what can be inferred from all this.
As anticipated in the introduction, the questionnaire in \cite{DBLP:conf/hhai/AlfieriDGGS23} is made of scenarios. A human agent provides feedback by answering to the questionnaire one scenario at a time. We report here 
two
% three 
of those scenarios that will be used in the paper.
\begin{scenario}
\label{scenario:postoffice}
As I am about to leave the post office, the queue-eliminating machine breaks down. A messy
line is forming, and a clerk starts hand-writing numbered cards for people coming in. Do I stop and
help him?
 Let us call this scenario \scen{postoffice}.
\end{scenario}
% \begin{scenario}
% \label{scenario:wallet}
% I am taking a walk and find a wallet with €1,000 inside. There is no ID in it. Do I turn it in to
% the nearby police station? 
%  Let us call this scenario \scen{wallet}.
% \end{scenario}
%\comment{We do not refer to this scenario, shall we remove it?}
\begin{scenario}
\label{scenario:fruits} There are trees with ripe fruit in a private park with private access. The gate is open and there are no people around. Do I go in and steal some? Let us call this scenario \scen{fruits}.
\end{scenario}
%%% INSERT PIC
% pics from original sketches and augmented with https://www.promeai.com/
%\documentclass{IOS-Book-Article}
%\usepackage[percent]{overpic}
%\usepackage{graphicx,tikz}
%\usepackage{amsmath}

%\begin{document}
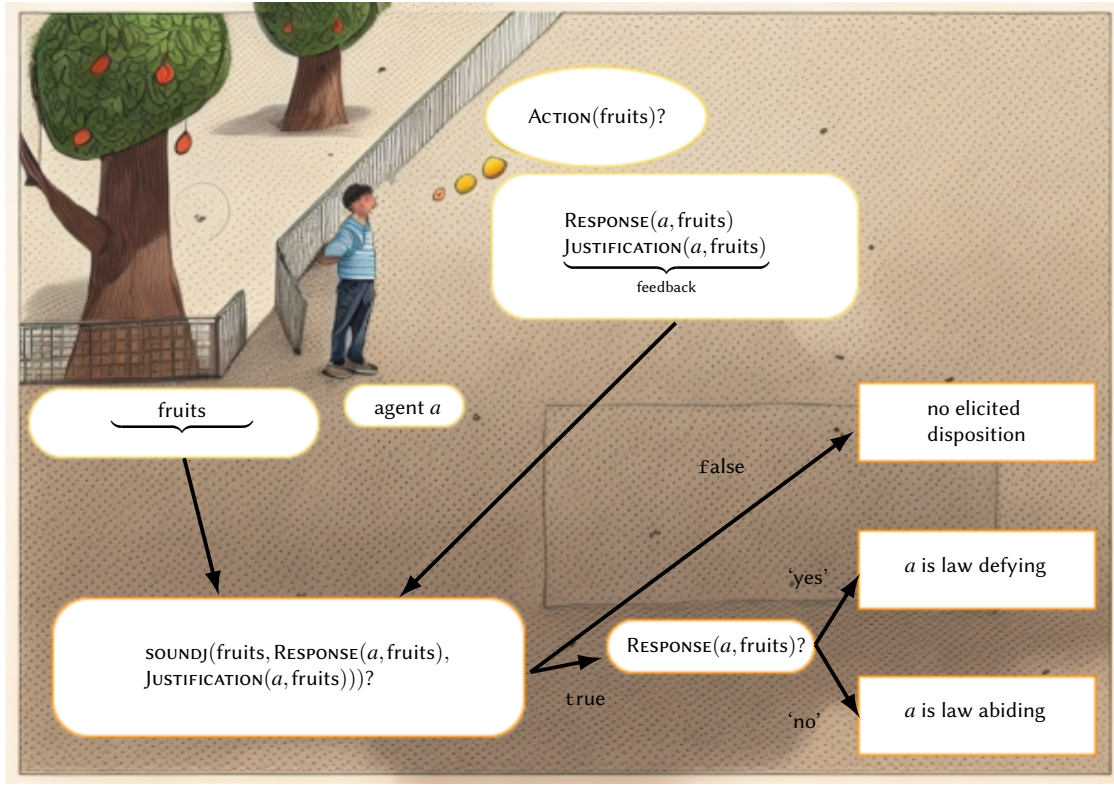
\begin{figure*}
    \centering
%\begin{overpic}[width=0.95\linewidth,grid,tics=10]{scenario1-base-edited.png}
%\scalebox{0.5}{
\begin{overpic}[width=1.0\linewidth]{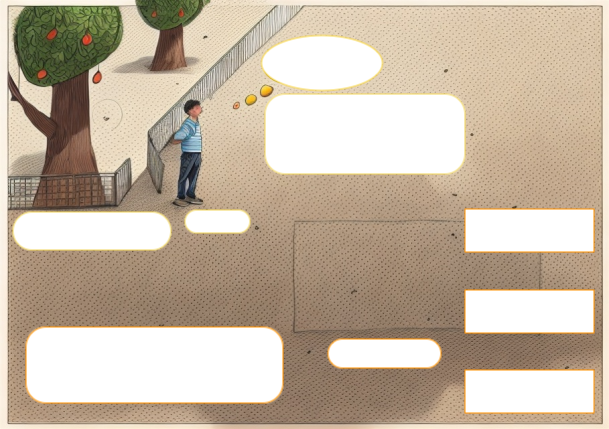}
 \put (33,33.3) {\small agent $a$}
 \put (46.8,59.5) {\small $\action(\scen{fruits})$?} % YES / NO}
 % \put (3,34) {\small  $\underbrace{\phantom{XXXXXXXXXXXXX}}_{\scen{fruits}) = \{\param 4\}}$}
  \put (9.5,33) {\small  $\underbrace{\phantom{XXX}\scen{fruits}\phantom{XXX}}$}
  
  \put (50,49) {$\underbrace{
  \begin{minipage}{2.7cm}\small
      $\response(a, \scen{fruits})$\\
      $\justification(a, \scen{fruits})$
   \end{minipage}}_{\text{feedback}}$}

\put(16,29.3){\linethickness{0.5mm} \vector(1,-4){3.1}} % from scenario PRESS

\put(60,41.5){\linethickness{0.5mm} \vector(-3,-3){24.8}} % from feedback
\put(12.5,10) {\begin{minipage}{3.2cm}\small$\sound(\scen{fruits},\response(a, \scen{fruits}),\linebreak\justification(a,\scen{fruits})))$?\end{minipage}}

 \put (47,10){\linethickness{0.5mm} \vector(5,1){6.5}}
 \put (50,7) {\texttt  true}

 \put (47,10){\linethickness{0.5mm} \vector(20,15){29}} 
 \put (62,28) {\texttt false}

\put (55.6,11.8) {\small
$\response(a,\scen{fruits})$?}

  \put (82.5,32) {\begin{minipage}{3.2cm}\small
no elicited\\ disposition
\end{minipage}}
  \put (80.5,19) {$a$ is law defying}
  \put (80.5,6) {$a$ is law abiding}

\put (72.5,12.5){\linethickness{0.5mm} \vector(2.5,4){4}} 
\put (70,18) {`yes'}

\put (72.5,12.5){\linethickness{0.5mm} \vector(2.4,-4){4}}
\put (70,5) {`no'}
\end{overpic}
%}
% ACKNOWLEDGMENT: https://www.promeai.com/
\caption{\label{fig:picture-fruits} From scenarios to soft ethics. Depiction of \Cref{scenario:fruits}, called $\scen{fruits}$, with some elements of formalisation, and commented in more details in \Cref{example:fruits-consistency}.}
\end{figure*}

%\end{document}

%%%%%%%%
After answering the questions of the scenarios by `yes' or `no', the subjects are asked to justify their answer by assigning values to four parameters. The parameters used in \cite{DBLP:conf/hhai/AlfieriDGGS23} are viewed as fundamental principles upon which ethics theories are usually constructed. The four parameters are as follows: 
\begin{description}
    \item \param 1 How much did the potential consequences of the action on others weigh on my choice?
    \item \param 2 How much did the potential consequences of the action on me weigh on my
choice?
    \item \param 3 How much did my personal experiences weigh on my choice?
    \item \param 4 How much did respect for the law weigh on my choice?
\end{description}
%\noindent These parameters isolate two meta-values. The first two parameters point out to other-regarding concerns while the last two point to self-regarding concerns.\co{CHECK THAT}

\medskip
Since we are interested in an overall representation of the soft-ethics preferences, we make a particular effort at extracting the concepts and the relations among them that are involved in the more informal presentation from \cite{DBLP:conf/hhai/AlfieriDGGS23}.

The first concept in the domain of the questionnaire is that of a \emph{human 
 agent} $a$. It is typically an individual decision maker from whom the soft ethics is revealed through their answers to the questionnaire. Then we have the scenario, which is the central concept. 
%\comment{I have changed context(s) into setting(s). The setting(s) is the first part of the scenario. The context, would be the whole scenario somehow.}
%
A \emph{scenario} $s$ is made of:
\begin{itemize}
\item A setting $\setting(s)$ which is a description in natural language of the setting of the scenario $s$.
\item A problem $\problem(s)$ which is a description in natural language of the problem of the scenario $s$.
\item An action $\action(s)$ which is a description in natural language of an hypothetical action that the human agent might perform of not.
\end{itemize}
\noindent The \emph{set of scenarios} is noted $\scenarios$.

%\comment{for later (?): ideally we should have the specification of the ``direction'' of the action, or expected ``good'' answer; the one that will elicit a disposition when soundly justified.}
%
\begin{example}
Consider \Cref{scenario:postoffice}. We have the setting $\setting(\scen{postoffice})$ which is ``As I am about to leave the post office, the queue-eliminating machine breaks down.'', the problem $\problem(\scen{postoffice})$ which is ``A messy line is forming and a clerk starts hand-writing numbered cards for people coming in.'', and the action $\action(\scen{postoffice})$ which is ``stop and help him''.

All the provided information can be interpreted as stimuli. That is, as properties that may trigger some disposition of the individual. $\setting(\scen{postoffice})$ provides the property of a state-based disposition ``readiness to leave'', and ``machine broken''.
$\problem(\scen{postoffice})$ provides the properties ``messy line forming'', and ``clerk hand-writes numbers and is needing help''.
$\action(\scen{postoffice})$ provides the action ``stop and help''.
\end{example}
The properties of a scenario, and the moral and behavioural properties of the agent are stimuli-disposition partners, as bearers of properties that may reveal themselves by interacting with each other. Once the agent is in the given setting, their dispositions, in the form of potential behaviours, are triggered by the properties of the overall scenario: the setting, the problem, and the action.

A scenario is qualified with the help of a \emph{set of parameters} \parameters.
As in the questionnaire of \cite{DBLP:conf/hhai/AlfieriDGGS23} which informs our study, we are interested in the social and ethical domain. The questionnaire
uses the set $\{\param 1, \param 2, \param 3, \param 4\}$ to justify the actions in the scenarios.
However, we prefer to reformulate the wording of the parameters. For, as we said in \Cref{sec:dispositions}, dispositions are gradable properties. For instance, \param 1 is about whether one is willing to help others. As an extreme case, one in which the subject assigns the top value ($5$) to the parameter, it reveals the user's altruistic disposition. Moreover, in order for an action to be altruistic, the action should be considered positive. So, to satisfy gradability each parameter must run along a determinate-determinable dimension.
(This is analogous to a physical object parameter running along a breakable--fragility dimension, where the `fragility' is determinate and  `breakability' is a determinable dimension.) With respect to parameter \param 1 that would be \emph{good willingness--altruism}; with respect to parameter \param 2 we propose \emph{self-servingness--egoism}; as to parameter \param 3 we propose \emph{pragmatism--expertness};
%\comment{changed expertiveness for expertness; not sure it's the right one yet. OK} 
finally, parameter \param 4 is \emph{legality--obedience}.
%\comment{changed strict for obedience OK} 
So, we may re-phrase the parameter by adding also the positivity of the action in the first two parameters, those that are other-regarding. By ``positive effect'' we notice that this should be from the point of view of the other human agents. Moreover, the action should not be taken for granted from all parties (so it is not obvious that the action is going to be performed) and, being positive means that they are desirable from other human agents' point of view.

Summing up:
\begin{description}
    \item{\param 1} refers to the human agent's consideration about the positive effect of their action on others.
    %``How much did the potential consequences of the action on others weigh on my choice?''
    It takes values on an interval scale of \emph{altruism} over a \emph{goodwill} dimension.
    
    \item{\param 2} refers to the human agent's consideration about the positive effect of their action on themselves: 
    %``How much did the potential consequences of the action on me weigh on my choice?''
    It takes values on an interval scale of \emph{egoism} over a \emph{self-servingness} dimension.
    
    \item{\param 3} refers to the human agent's consideration about their personal experiences: 
    %``How much did my personal experiences weigh on my choice?''
    It takes values on an interval scale of \emph{expertness} over a \emph{pragmatism} dimension.
    
    \item{\param 4} refers to the human agent's consideration about the law: 
    %``How much did respect for the law weigh on my choice?''
    It takes values on an interval scale of \emph{obedience} over a \emph{legality} dimension.
    
\end{description}
It should be further noticed that a scenario $s$ may stress one or more of the parameters in \parameters.
The set $\press(s) \subseteq \parameters$ is the set of parameters that the scenario $s$ puts pressure on. 
\begin{example}
Consider \Cref{scenario:postoffice}. The scenario $\scen{postoffice}$ presses the parameter about the consequences of the action on others. Hence, $\press(\scen{postoffice}) = \{\param 1\}$.
\end{example}

This is typically intended and determined by the designer of the scenario. But this could also be determined experimentally if need be.

The human agents' feedback on a scenario uses an interval scale from $1$ to $5$: $\scale = \{1,2,3,4,5\}$.

A \emph{feedback} $f$ on a scenario $s$ provided by a human agent $a$ is made of:
\begin{itemize}
    \item a response $\response(a,s) \in \{yes, no\}$ indicating whether the human agent $a$ would perform action $\action(s)$ if confronted to scenario $s$.
    \item a justification $\justification(a,s) \in \scale^\parameters$, where the integer value\linebreak $\justification(a,s)(\param i)$ indicates the level of relevance of \param i for human agent $a$ in choosing $\response(a,s)$.\footnote{We use the standard notation where $X^Y$ denotes the set of functions from set $Y$ to set $X$.}
\end{itemize}

\begin{figure}
%\begin{wrapfigure}{R}{0.45\textwidth}
    \centering
    \usetikzlibrary{positioning,shapes.geometric}

% For drawing
\def\firstellip{(1.6, 0) ellipse [x radius=3cm, y radius=1.5cm, rotate=50]}
\def\secondellip{(0.3, 1cm) ellipse [x radius=3cm, y radius=1.5cm, rotate=50]}
\def\thirdellip{(-1.6, 0) ellipse [x radius=3cm, y radius=1.5cm, rotate=-50]} 
\def\fourthellip{(-0.3, 1cm) ellipse [x radius=3cm, y radius=1.5cm, rotate=-50]}
\def\bounding{(0, 1cm) ellipse [x radius=6cm, y radius=4.5cm, rotate=0]}

%def\firstcircle {(0,0)    circle (3 cm)}
%\def\secondcircle{(45:2cm) circle (3 cm)}
%\def\thirdcircle {(0:3cm)  circle (3 cm)}
%\def\fourthcircle{(-45:2cm)circle (3 cm)}
\begin{tikzpicture}[scale=0.45]
%\draw \firstcircle node[below] {$$};
%\draw \secondcircle node [above] {$$};
%\draw \thirdcircle node [below] {$$};
%\draw \fourthcircle node [below] {$$};
\begin{scope}[shift={(0cm,0cm)}, fill opacity=0.4]
    \fill[blue]            \firstellip;
    \fill[orange] \secondellip;
    \fill[green]  \thirdellip;
    \fill[red]\fourthellip;
\end{scope}          % <-- Move the \end{scope} to here for clear labels.
%   \draw \firstcircle  node[below] {$$};
%   \draw \secondcircle node [above] {$$};
%   \draw \thirdcircle  node [below] {$$};
%   \draw\fourthcircle  node[below]{$$};

    \draw \firstellip node [label={[xshift=0.9cm, yshift=-0.9cm]\param 1}] {};
    \draw \secondellip node [label={[xshift=1.3cm, yshift=0.7cm]\param 2}] {};
    \draw \fourthellip node [label={[xshift=-1.3cm, yshift=0.7cm]\param 3}] {};
    \draw \thirdellip node [label={[xshift=-0.9cm, yshift=-0.9cm]\param 4}] {};
    \draw \bounding node
    [label={[xshift=3cm, yshift=1.3cm]\scenarios}] {};
    %[label=above left:\scenarios] {};

    \node at (2.5,0.2) [circle,fill,inner sep=1.5pt,label=\scen{postoffice}]{};
    \node at (-2.4,-0.8) [circle,fill,inner sep=1.5pt,label=\scen{fruits}]{};
    
 %\end{scope}          % <-- move this line up will get a clear labels for circle names. It has opacity=0.4 currently.
 \begin{scope}
   \clip \firstellip;
   \clip \secondellip;
   \clip \thirdellip;
   %\fill[lightgray]
   \clip \fourthellip;
 \end{scope}
  \draw \firstellip;
  \draw \secondellip;
  \draw \thirdellip;
  \draw \fourthellip;
 % \node[draw,text width=2.5cm] at (1.5,0) {some text spanning three lines with automatic line breaks};
 \end{tikzpicture}
    \caption{\label{fig:enter-label} Categories of scenarios.}
%\end{wrapfigure}
\end{figure}
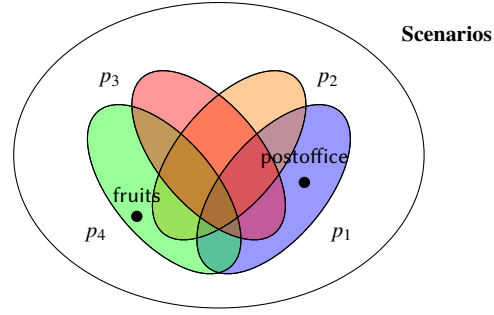

The \emph{category} $\category(s) \subseteq \scenarios$ of scenario $s$ is the set of scenarios $s'$ such that $\press(s) = \press(s')$. 
%
%With $\parameters = \{\param 1, \param 2, \param 3, \param 4\}$,
The 16 categories of scenarios can be visualised with the Venn's 4-set diagram represented on Figure~\ref{fig:enter-label}.

Each category is thus intended as an abstraction of a scenario. A soft-ethics preference elicited from a scenario is intended to apply to all scenarios belonging to the same category.
This is the primary mechanism to handle 
new situations encountered by human agents.
Elicited dispositions are then to be implemented into an ethical software profile that augments human capabilities by preserving their soft-ethical preferences in interactions with autonomous systems.

\section{A moral oracle}
\label{sec:consistency}

Before we can elicit a dispositional soft-ethics preference from a feedback, 
we will eventually need a mechanism to decide whether a scenario and a feedback follow a certain consistency. For now, we treat this mechanism as an oracle, $\sound$, that stands for ``sound justification''.
The difficulty resides in analysing formally the scenarios as described in \cite{DBLP:conf/hhai/AlfieriDGGS23}, and the 'direction' of the actions. E.g., in \Cref{scenario:postoffice}, an answer `yes' has a positive overtone, while in \Cref{scenario:fruits}, answer `no' has a negative overtone.
We only start specifying what this mechanism should do.

Let us consider a scenario $s$, that presses on the parameters $\press(s)$, and includes the action $\action(s)$ that can or cannot be performed. Let us also consider a human agent $a$ and $a$'s feedback that includes the answer yes or no $\response(s)$ and the justification $\justification(a,s)$ in terms of parameters values. 
Remember that if $\response(a,s)$ is yes, the agent takes action $\action(a)$, and
if $\response(a,s)$ is no, the agent does \emph{not} take action $\action(a)$.

We can define the boolean function
$\sound(s,\response(a,s), \justification(a,s)))$ which captures the judgement about whether the justification is sound with respect to the action taken in the scenario $s$ by the human agent $a$.
% This function guarantees that for any other human agent $a'$ for the same scenario $s$ $\justification(a',s)(\param i)$ such that $\param i \in \press(s)$ is the same/similar.

\begin{example}
For example let us consider again $\scen{postoffice}$ from \Cref{scenario:postoffice} (post office). Remember that the scenario presses on \param 1, that is, good-willingness.
Let us assume that:\footnote{The placeholder value $\_$ indicates that the exact value does not matter. We suppose that the oracle takes $1$ as a low value and $4$ as a high value.}
\begin{itemize}
    \item agent $a$ helps the clerk ($\response(a,s) = `yes'$) with justification $(4,\_,\_,\_)$,
    \item agent $b$ does not help the clerk ($\response(b,s) = `no'$) with justification $(1,\_,\_,\_)$,
    \item agent $c$ does help the clerk ($\response(c,s) = `yes'$) with justification $(1,\_,\_,\_)$,
    \item agent $d$ does not help the clerk ($\response(d,s) = `no'$) with justification $(4,\_,\_,\_)$.
\end{itemize}
Then 
\begin{itemize}
\item $\sound(s,\response(a,s),\justification(a,s))$ is true,
\item $\sound(s,\response(b,s),\justification(b,s))$ is true,
\item $\sound(s,\response(c,s),\justification(c,s))$ is false,
\item $\sound(s,\response(d,s),\justification(d,s))$ is false.
\end{itemize}
\end{example}
In the previous example, a `yes' answer has an ethically `positive´ connotation. This is in contrast with the next example.
\begin{example}\label{example:fruits-consistency}
    Let us consider \scen{fruits} \Cref{scenario:fruits}, also depicted on \Cref{fig:picture-fruits}. Agent $a$ is considering entering the private park and stealing a fruit. The scenario presses on parameter \param 4, that is, the legality of the action.
    If $\response(a, \scen{fruits})$ is `yes' and $\justification(a, \scen{fruits})$ gives a high value to \param 4, $\sound(\ldots)$ is false, and then we cannot elicit any disposition.
    Instead, if $\response(a, \scen{fruits})$ is `yes' and $\justification(a, \scen{fruits})$ gives a low value to \param 4, $\sound(\ldots)$ is true, and since $\response(a, \scen{fruits})$ is `yes', we can elicit the disposition of agent $a$ to be law defying.
\end{example}
The $\sound$ function thus occupies an instrumental role in our soft-ethics preferences from questionnaire feedback.
Before eliciting a disposition, $\sound(s,\response(x,s),\linebreak \justification(x,s))$ filters out the responses by a human agent $x$ that are not consistent with the intended meaning of the scenario $s$. 

For the time being, we assume the existence and computability of this function. As it may appear clear, the actual implementation of the function must account for a nuanced setting of the parameters, and some information about the `direction´ of the action in a scenario. We discuss future work related to the function $\sound$ in the next section.

\section{Outlook}
\label{sec:outlook}

%\paragraph{Where we stand.}
We clarified the ontology of the questionnaire of \cite{DBLP:conf/hhai/AlfieriDGGS23}. Guided by a pre-formalisation, we have also proposed how the empirical data collected through this questionnaire permits to elicit the feedback from the subjects into soft-ethics preferences. To this end, we have adopted a behavioural approach. Furthermore, we have argued for a dispositional perspective of these soft-ethics preferences.

The work done so far has permitted us to identify the necessary pieces of information present in a scenario and in the feedback to derive a soft-ethics preference. Nonetheless, we found a stumbling block, inasmuch that those are not sufficient. We have indeed recourse to an oracle to inform us about the soundness of the feedback with a given scenario. This is the first natural course of action for future work.

\paragraph{Future work.}
We plan to work on an concrete implementation of $\sound$ function.
Working with existing questionnaires, we will need methods to extract the relevant pieces of information from scenario written in natural language. This includes understanding the `direction' of the action, whether either a `yes' or `no' should be considered a `positive' action.

We also envisage that missing information from the questionnaire could be easily filled in by the designers. In a future iteration of this work, we anticipate making recommendations on how to design a questionnaire with additional data. This would enable the fully automated elicitation of feedback for soft-ethics preferences.

Another perspective for future work lies in developing a formal language to represent the soft-ethics preferences elicited from such a questionnaire. 
It could be an adaptation of so-called SLEEC rules \cite{DBLP:journals/mima/TownsendPANCCHT22} to personal ethics, formalised along the ideas presented in \cite{DBLP:conf/aaai/TroquardSIPS24}.
We anticipate that classical logic might be too coarse to capture their dispositional nature. Instead, we will explore the use of probabilistic rules or fuzzy logic \cite{10.1007/978-3-642-83590-2_9}.

Finally, we want to use the gathered preferences as dispositions to create a software profile that enhances human abilities by respecting their ethical choices when they interact with autonomous systems.

\bibliography{biblio}

\end{document}